\documentclass[10pt,prd,twocolumn]{revtex4-1}
\usepackage{hyperref,graphicx,color,amsmath, amssymb,natbib, url}

\def\jcap{J. Cosmol. Astropart. Phys.}%
\def\mnras{Mon.~Not.~R.~Astron.~Soc.}%
\def\prd{Phys.~Rev.~D}%
\def\prl{Phys.~Rev.~Lett.}%
\def\aap{A\&A}%

\def\figwidth{0.45\textwidth}
\begin{document}
\title{Observational effects of a running Planck mass}
\author{Zhiqi Huang$^{1}$}
\affiliation{${}^1$Canadian Institute for Theoretical Astrophysics, University of Toronto, 60 St. George Street, Toronto, Ontario, M5S 3H8, Canada}

\date{\today}

\begin{abstract}
  
  We consider observational effects of a running effective Planck mass in the scalar-tensor gravity theory. At the background level, an increasing effective Planck mass allows a larger Hubble constant $H_0$, which is more compatible with the local direct measurements. At the perturbative level, for cosmic microwave background (CMB) anisotropies, an increasing effective Planck mass (i) suppresses the unlensed CMB power at $\ell \lesssim 30$ via the integrated Sachs-Wolfe  effect and (ii) enhances CMB lensing power. Both effects slightly relax the tension between the current CMB data from the Planck satellite and the standard $\Lambda$CDM model predictions. However, these impacts on the CMB secondary anisotropies are subdominant, and the overall constraints are driven by the background measurements. Combining CMB data from the Planck satellite and an $H_0$ prior from Riess {\it et al}, we find a $\sim 2\sigma$ hint of a positive running of the effective Planck mass. However, the hint goes away when we add other low-redshift observational data including type Ia supernovae, baryon acoustic oscillations and an  estimation of the age of the Universe  using the old stars.
  
\end{abstract}

\maketitle

\section{Introduction}

The recent observations of cosmic microwave background (CMB) temperature and polarization anisotropies by the Planck satellite~\cite{Planck2015OverView, Planck2015Params}, once again, have confirmed the standard $\Lambda$CDM model, that is, a Universe with a cosmological constant $\Lambda$,  cold dark matter (CDM), and initial metric fluctuations seeded by an early inflationary phase.

While no strong evidence against the concordance model has been found, however,  some $\sim 2\sigma$ hints of deviations from $\Lambda$CDM model have emerged. On large angular scales (multipole $\ell \lesssim 40$), CMB temperature power tends to be lower than the $\Lambda$CDM-model prediction~\cite{Planck2015IandS}. On small angular scales, a $22\% \pm 10\%$ excess of lensing potential power spectrum, the so-called $A_L$ anomaly, has been found in the Planck temperature data~\cite{Planck2015Params}. Finally, the locally measured Hubble constant $H_0$ appears to be higher than Planckian $\Lambda$CDM prediction~\cite{HSTRiess, Verde:2013wza, Amendola15, Umilta15,Ballardini16}. Ref.~\cite{HSTEfstathiou} revisited this problem and found that the locally measured $H_0$ value is lower if a revised geometric maser distance to NGC 4258 is used. This may suggest some unknown systematics biasing the local $H_0$ measurement. In this article, however, we assume that the $H_0$ tension is not due to unknown systematics and only study it from the theoretical aspect.

Phenomenologically modified gravity as an alternative to the cosmological constant has been proposed as a possible solution to the lensing power excess~\cite{Planck2015DE, Valentino15}. In this article, we proceed in this direction by considering a concrete model with a running effective Planck mass. The planck mass run can be described self-consistently in the the Horndeski gravity theory, which is the most general scalar-tensor gravity theory with at most second derivatives in the equation of motion. Linear perturbations of Horndeski gravity can be described in the effective field theory (EFT) language with five background functions~\cite{Creminelli09, Gleyzes13, Gubitosi13, Bellini14, UDE}: the dark energy equation of state $w$, the effective Planck mass run rate $\alpha_M$, the tensor speed excess $\alpha_T$, the kineticity $\alpha_K$, and the braiding $\alpha_B$. For simplicity of discussion, we define $\alpha_W = 1 + w$, such that for the $\Lambda$CDM model, all the $\alpha$ functions are zero.

From a theoretical aspect, all the $\alpha$ functions can be nonzero, and we might lose the power of predictability with current limited data at low redshift and so many extra degrees of freedom in the dark energy theory. While the future dark energy surveys may change the situation~\cite{Gleyzes15}, in this work we will only consider current observations and limit the degrees of freedom by studying a very special case where only the effective Planck mass run rate $\alpha_M$ is allowed to be nonzero. In this ``$\alpha_M$CDM'' model, the present-day gravity may be weaker or stronger than that during recombination. We assume a chameleon screening mechanism~\cite{Chameleon1, Chameleon2} where the effective Planck mass $M_*$ in high matter density regions equals the bare Planck mass $M_p = 1/\sqrt{8\pi G_N}$, $G_N$ being the Newton's gravitational constant.

The EFT dark energy model discussed in Ref.~\cite{Planck2015DE} differs from the $\alpha_M$CDM cosmology in two aspects: ({\it i}) in Ref.~\cite{Planck2015DE} $\alpha_B$ and $\alpha_K$ are derived from the assumption that the spatial derivatives are limited to at most second order, too, whereas in $\alpha_M$CDM model they are both fixed to be zero, and ({\it ii}) in Ref.~\cite{Planck2015DE} the background expansion history is fixed to be identical to the $\Lambda$CDM model, whereas in the $\alpha_M$CDM model, we fix the Jordan-frame dark energy equation of state, as defined in, e.g., Eq.~(2.28) of Ref.~\cite{Gleyzes15}, to be $-1$. 

This paper is organized as follows. In Sec.~\ref{sec:model} we describe the main differences between $\alpha_M$CDM model and $\Lambda$CDM model from the theoretical point of view. In Sec.~\ref{sec:ano}, we compute the observational effects and discuss the above-mentioned anomalies in the context of $\alpha_M$CDM cosmology. Sec.~\ref{sec:con} concludes.

\section{$\alpha_M$CDM Model \label{sec:model}}

Assuming a flat universe, the Friedmann-Lemaitre-Robertson-Walker metric including scalar perturbations can be written in the Newtonian gauge as
\begin{equation}
ds^2 = -(1 + 2\Phi)d t ^2 + (1-2\Psi) a^2(t) \mathbf{dx}^2 \;,
\end{equation}
where $a(t)$ is the scale factor normalized to $a=1$ today, $t$ the cosmological time, $\Phi$ the Newtonian potential, and $\Psi$ the space curvature. The expansion rate $H$ is defined as $H = {\dot a}/ a$, where a dot denotes the derivative with respect to $t$. The current $H$ value is the so-called Hubble constant $H_0$, where the subscript $0$ denotes its value today.

The effective Planck mass run rate is defined as
\begin{equation}
  \alpha_M(a) \equiv \frac{d\ln \left(M_*^2\right)}{d\ln a} \,,
\end{equation}
where $M_*$ is the effective Planck mass in the Jordan frame. The background function $\alpha_M(a)$ is obtained by perturbatively expanding the Horndeski gravity action to second order~(see, e.g.,~\cite{UDE}). 

Since GR is extremely successful at describing the primary CMB fluctuations, we do not want to modify the background and perturbation equations in the early Universe where the dark energy contribution to the total energy density is negligible. A natural choice of the functional form of $\alpha_M(a)$, which is also used in Refs.~\cite{Planck2015DE, Gleyzes15}, is then
\begin{equation}
  \alpha_M(a) = \alpha_{M0} \frac{\Omega_{\rm DE}(a)}{\Omega_{\rm DE,0}} \, , \label{eq:alphaM}
\end{equation}
where $\Omega_{\rm DE}$ is the energy fraction of dark energy and $\Omega_{\rm DE,0}$ the present value of $\Omega_{\rm DE}$. To simplify the calculation, instead of using $\Omega_{\rm DE}$ of the $\alpha_M$CDM model, in Eq.~\eqref{eq:alphaM} we use $\Omega_{\rm DE}$ from a fiducial $\Lambda$CDM model with fixed $\Omega_{DE,0} = 0.7$.

In $\alpha_M$CDM cosmology, the Friedmann equations become
\begin{equation}
  \frac{d (H^2)}{d\ln a} = - \frac{\rho_{\rm tot} + p_{\rm tot}}{M_*^2}\, , \label{eq:f1}
\end{equation}
where $\rho_{\rm tot}$ and $p_{\rm tot}$ are total density and total pressure of all species including dark energy, and 
\begin{equation}
  H^2 = \frac{\rho_{\rm tot}}{3M_*^2}\, . \label{eq:f2}
\end{equation}

The perturbation equations are lengthy, and we refer the reader to Ref.~\cite{UDE}. Here we only summarize the key difference between the $\alpha_M$CDM model and  the $\Lambda$CDM model.

In the late Universe where the anisotropy of radiation and neutrinos can be ignored, the gravitational potential $\Phi$ deviates from the space curvature in the $\alpha_M$CDM model:
\begin{equation}
  \Phi = (1 + \alpha_M)\Psi \, . \label{eq:aniso}
\end{equation}
The Poisson equation for $\alpha_M$CDM cosmology reads
\begin{equation}
  \frac{k^2}{a^2} \Phi = -\frac{\rho_m(1+\alpha_M)}{2M_*^2} \delta_m \, . \label{eq:poisson}
\end{equation}
where $k$ is the comoving wave number, $\rho_m$ the background matter energy density, and $\delta_m$ the matter density fluctuation in a comoving synchronous gauge.

We parametrize $\alpha_M$CDM cosmology with the following parameters: ({\it i}) the present-day effective Planck mass run rate $\alpha_{M0}$ , ({\it ii}) the physical baryon density $\Omega_b h^2 (M_{*0}/M_p)^2$, ({\it iii}) the physical CDM density $\Omega_c h^2 (M_{*0}/M_p)^2$, {\it iv}) the angular extension of the sound horizon on the last scattering surface $\theta$, ({\it v}) the reionization optical depth $\tau$, and for the primordial power spectrum of scalar perturbations $\mathcal{P}_s(k) = A_s (k/k_{\rm p})^{n_s-1}$ ($k_{\rm p} = 0.05\,\mathrm{Mpc}^{-1}$), we use ({\it vi}) $n_s$ and ({\it vii}) $\ln ( A_se^{-2\tau})$. For the primordial helium abundance and the sum of neutrino masses, we fix them to be $Y_p = 0.248$ and $\sum m_{\nu}=0$. We assume the tensor-to-scalar ratio $r$ is negligible and ignore primordial gravitational waves.

The advantage of using these parameters is that the primary CMB power spectrum is $\alpha_{M0}$ independent. The impact of $\alpha_{M0}$ on the CMB power spectrum is only through the integrated Sachs-Wolfe (ISW) effect and the lensing contribution. (Other secondary effects are treated as foreground templates.) 

To solve the background evolution and linear perturbations, we implement Eqs.~(109)-(117) in Ref.~\cite{UDE} with a modernized version of the package COSMOLIB~\cite{CosmoLib}. The new code, which has been upgraded with FORTRAN 2008 object-oriented features, is renamed as the COSMOLOGY OBJECT-ORIENTED PACKAGE (COOP). COOP is publicly available at \url{http://www.cita.utoronto.ca/~zqhuang/coop}\,. For parameter constraints, we use a COOP built-in Monte-Carlo Markov chain (MCMC) simulator.

\section{Impact on the Anomalies} \label{sec:ano}

\subsection{Background evolution and the $H_0$ anomaly}

For fixed $\Omega_b h^2(M_{*0}/M_p)^2$, $\Omega_c h^2(M_{*0}/M_p)^2$, and $\theta$, the parameter $\alpha_{M0}$ is degenerate with $H_0$. If only primary CMB anisotropies were measured, this degeneracy would be perfect. In practice, ISW and lensing effects slightly break the degeneracy. Moreover, low-redshift surveys can probe the geometry of the late Universe and further break the degeneracy. Here we consider baryon acoustic oscillations (BAO)~\cite{BAOSDSS, BAOBOSS, BAO6DF}, type Ia supernovae (SN)~\cite{SNJLA}, and a conservative Gaussian prior on the age of the Universe ($14.4\pm 0.7$ Gyr) from estimation of the age of the old stars~\cite{Verde:2013wza}. For $H_0$ we use a Gaussian prior $H_0 = 73.8 \pm 2.4\,\mathrm{km}\,\mathrm{s}^{-1}\,\mathrm{Mpc}^{-1}$ \cite{HSTRiess}. See Ref.~\cite{Planck2015Params} for a more detailed description of these data sets.

The CMB constraints on the background parameters $\Omega_b h^2(M_{*0}/M_p)^2$, $\Omega_c h^2(M_{*0}/M_p)^2$ and $\theta$ are not sensitive to $\alpha_{M0}$ which only has secondary effects on CMB. Thus, we can define a compressed CMB likelihood as an approximation of ``background-only'' constraint:
\begin{equation}
  \ln \mathcal{L}_{\rm BG} = -\frac{1}{2} \left(\mathbf{v} - \bar{\mathbf{v}}\right) C^{-1} \left(\mathbf{v} - \bar{\mathbf{v}}\right)^T \, ,\label{eq:likecmb}
\end{equation}
where $\mathbf{v} \equiv (\Omega_b h^2(M_{*0}/M_p)^2, \Omega_c h^2(M_{*0}/M_p)^2, \theta)$ is the parameter vector, $\bar{v}$ and $C$ its mean and covariance matrix for the $\Lambda$CDM model and Planck temperature and the low-$\ell$ polarization  data (TT + lowP).

\begin{figure}
  \includegraphics[width=\figwidth]{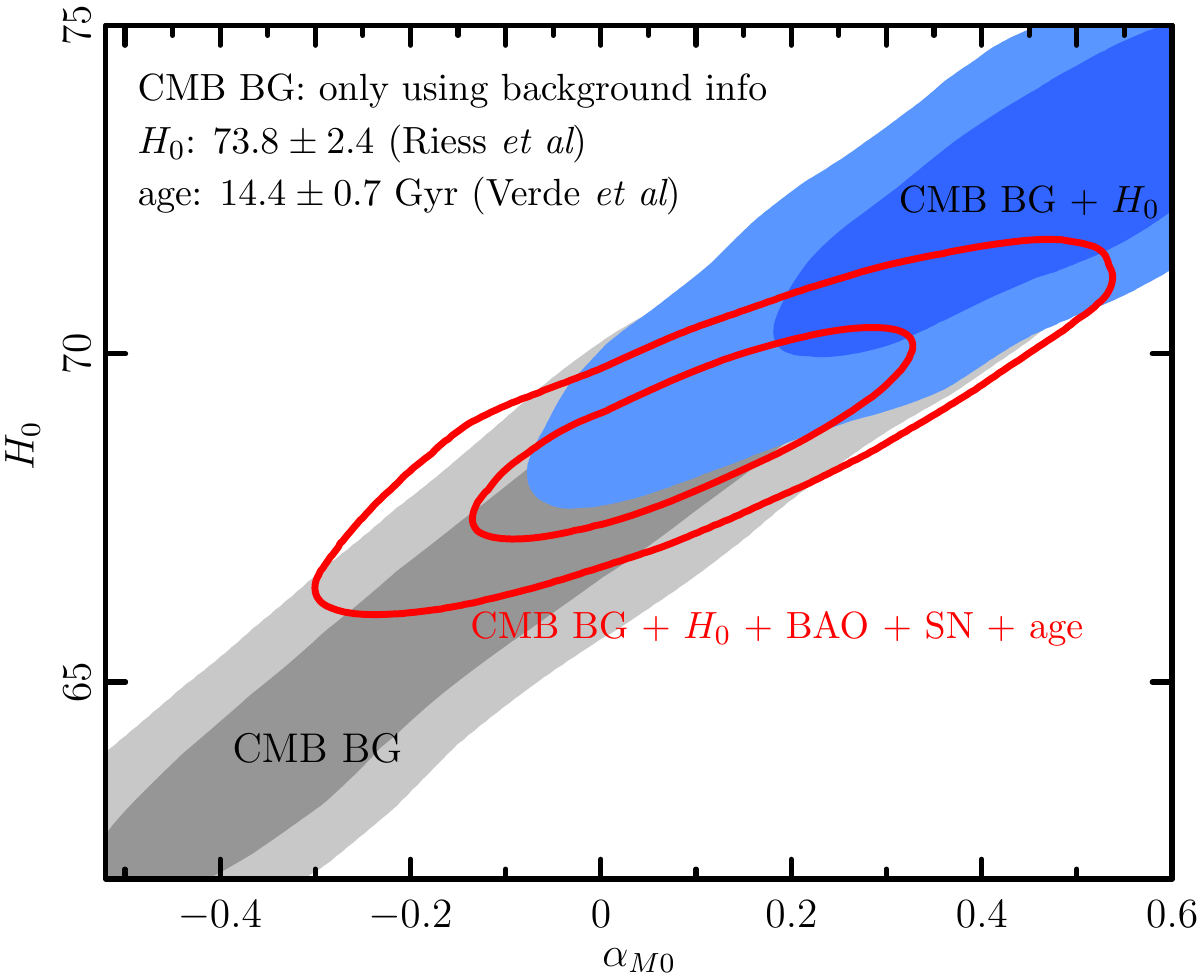}
  \caption{Marginalized constraint on $H_0$ and $\alpha_{M0}$ with compressed CMB likelihood eq.~\eqref{eq:likecmb}, SN and BAO likelihoods, age prior, and $H_0$ prior. Contours are 68.3\% C.L. and 95.4\% C.L.. \label{fig:H0alphaM0}}
\end{figure}

With the CMB ``background-only'' likelihood Eq.~\eqref{eq:likecmb}, SN and BAO likelihoods, the age prior, and the $H_0$ prior, we obtain a joint constraint on $H_0$ and $\alpha_{M0}$ shown in Fig~\ref{fig:H0alphaM0}. When only the CMB ``background-only'' constraint is used, $\alpha_{M0}$ and $H_0$ are perfectly degenerate. The addition of $H_0$ prior pushes $H_0$ to a higher value and disfavors $\Lambda$CDM ($\alpha_M=0$) at $\sim 2\sigma$ level. With all the other data sets (BAO, SN and age prior) included, however, the $\Lambda$CDM model falls back into the $1\sigma$ contour.

The positive correlation between $H_0$ and $\alpha_{M0}$ can be understood analytically. With the CMB constraint, we can approximate that the initial conditions are fixed at recombination. Given $\alpha_W = 0$ the dark energy contribution to the numerator of the right-hand side of Eq.~\eqref{eq:f1} vanishes. The matter contribution to the numerator is $\alpha_M$-independent, too. A positive $\alpha_{M0}$ leads to a larger denominator $M_*^2$ at late time, which makes $H^2$ drop less rapidly as the Universe expands.

\subsection{Linear perturbations and the impact on CMB}

In this subsection, we consider the impact of $\alpha_M$ on the secondary CMB anisotropies (ISW and lensing) that we ignored in the last subsection.

\begin{figure}
  \includegraphics[width=\figwidth]{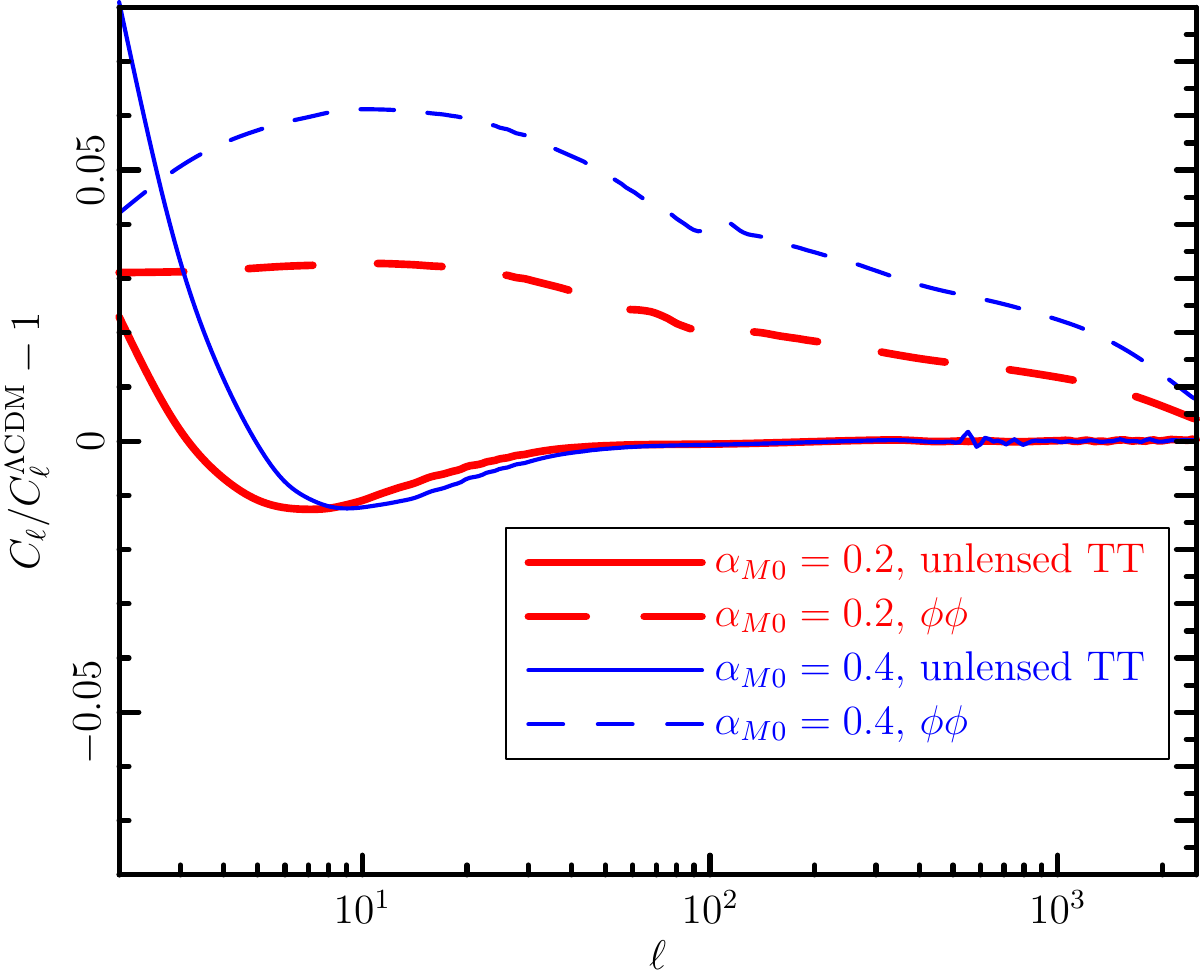}
  \caption{Relative difference between $\alpha_M$CDM model and $\Lambda$CDM model. The solid lines are unlensed CMB temperature power spectrum $C_{\ell}^{TT}$.  The dashed lines are lensing potential power spectrum $C_\ell^{\phi\phi}$. Thick and thin lines correspond to $\alpha_{M0} = 0.2$ and $\alpha_{M0} = 0.4$, respectively. \label{fig:cl}}
\end{figure}

\begin{figure}
  \includegraphics[width=\figwidth]{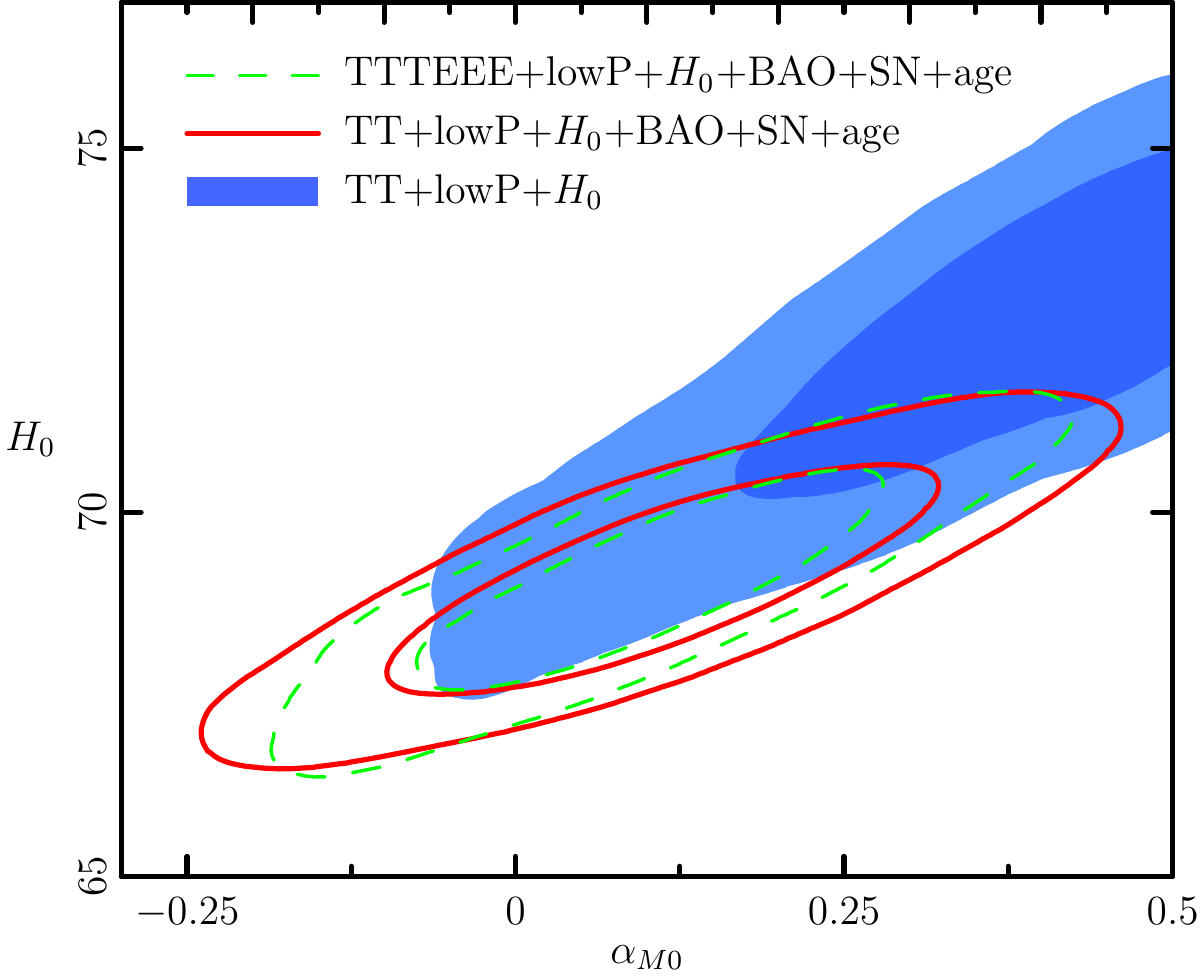}
  \caption{Marginalized constraint on $H_0$ and $\alpha_{M0}$ with full Planck likelihood, SN and BAO likelihoods, age prior, and $H_0$ prior. Contours are 68.3\% C.L. and 95.4\% C.L.. \label{fig:fullmcmc}}
\end{figure}

Before doing a full calculation, it is useful to qualitatively understand how strong these effects are. We fix the six standard parameters, i.e., the parameters excluding $\alpha_{M0}$,  to the $\Lambda$CDM best-fitting values and compute CMB power spectrum for various values of $\alpha_{M0}$. The results, shown in Fig~\ref{fig:cl}, indicate that the impact of an $\alpha_M\sim 0.1$ on the secondary CMB anisotropies is rather small compared to the Planck sensitivity, albeit in the ``right direction'' to relax the  $A_L$ tension and to reduce the low-$\ell$ power deficit. We, therefore, expect that the additional constraint from CMB secondary anisotropies does not make a big difference than the ``background-only'' case. This is confirmed by a full MCMC calculation using the publicly available Planck likelihood for temperature and low-$\ell$ polarization data  (TT + lowP)~\cite{Planck2015Like}. The results are shown in Fig~\ref{fig:fullmcmc}. For comparison we also show the result with high-$\ell$ polarization data included (TTTEEE + lowP), which is not very different from the TT + lowP case or the ``background only'' case.

So far we have fixed $\alpha_K = \alpha_B = 0$. If the full theory is dynamical and weakly coupled, its time kinetic term (proportional to $\alpha_K + 6\alpha_B^2$~\cite{UDE}) in general will not be perfectly zero. For $\alpha_M \ge 0 $ this is not a problem, we can interpret our calculation as an approximation for $0 \le \alpha_K \ll 1$. Indeed, for $\alpha_M\ge 0$ and $ \alpha_K \ge 0$, we find the CMB power spectrum continuously depend on $\alpha_K$ and a small $\alpha_K<0.1$ has negligible observational effect with current data sets ($\delta\chi^2 \ll 1$). For $\alpha_M<0$, however, caution needs to be taken: a slightly tuned up $\alpha_K$ or $\alpha_B$ will ruin the stability of the perturbation equations and lead to exponential growth of dark energy fluctuations. The reader should bear in mind that the parameter space with $\alpha_M<0$ and all other $\alpha$ functions vanishing, although allowed by the data and healthy at quadratic level, may be difficult to construct in a full theory.

\section{Conclusions and Discussions \label{sec:con}}

In this article we have studied a concrete modified gravity model, where an non-vanishing anisotropy stress appears in the late Universe and the Poisson equation is modified {\it in a self-consistent way}. In this model, both the background evolution and the linear perturbations depend on the effective Planck mass run rate $\alpha_M$. The modified linear perturbations do have impacts on CMB secondary anisotropies; however, it is typically too small to be measured. 

Combined CMB + $H_0$ prior give a $\sim 2\sigma$ hint of a positive $\alpha_{M0}$, which is mostly driven by the $H_0$ tension at the background level. This hint goes away when other low-redshift background constraints (BAO + SN + age prior) are added.

At the background level, the $\alpha_{M}$CDM model is degenerate with a $w$CDM model where the dark energy equation of state deviates from $-1$. To distinguish them, we need to consider the late Universe cosmological perturbations that, in principle,  can be either  directly measured with redshift surveys or indirectly constrained by  CMB secondary anisotropies. In this work we have shown that CMB secondary anisotropies are not sensitive to an $\alpha_{M0} \sim 0.1$. (Larger $\alpha_{M0}$'s are ruled out by background constraints). We have not discussed the redshift survey data such as cluster abundance and weak-lensing measurements. Both of them involve modeling of halo mass function, which in $\alpha_M$CDM cosmology may be very different from the $\Lambda$CDM case. We will leave the study of halo mass function in $\alpha_M$CDM cosmology for a future work.

\section*{Acknowledgements}

We thank Filippo Vernizzi for very useful discussions and helps on developing the EFT dark energy code in COOP.

\end{document}